\documentclass[english,aps, superscriptaddress, floatfix, 12pt]{revtex4}
\usepackage[T1]{fontenc}
\usepackage[latin1]{inputenc}
\usepackage{graphicx}
\usepackage{amssymb}
\usepackage{epsfig}
\usepackage{amsmath}
\usepackage{psfrag}

\usepackage{babel}
\makeatother
\begin{document}

\title{Pressure and interaction measure of the gluon plasma}

\author{D. Antonov\\
{\it Fakult\"at f\"ur Physik, Universit\"at Bielefeld, D-33501 Bielefeld, Germany}\\
H.-J. Pirner, M.G. Schmidt\\
{\it Institut f\"ur Theoretische Physik, Universit\"at Heidelberg,}\\
{\it Philosophenweg 16} \& {\it 19, D-69120 Heidelberg, Germany}}

\begin{abstract}
We explore the thermodynamics of the gluon plasma in SU(3) Yang-Mills
theory emerging from the non-trivial spatial dynamics of valence gluons. 
The lattice data suggest that these gluons interact with each other linearly at large spatial separations. At high temperatures, valence gluons should reproduce the pressure of the 
non-interacting Stefan-Boltzmann plasma along with the leading perturbative correction.
These properties of valence gluons can be modeled in terms of the integral over their trajectories. 
We calculate such a world-line integral analytically and obtain the pressure and the interaction measure
$(\varepsilon-3p)/T^4$ of the gluon plasma. Additionally, we account for the contributions of 
stochastic background fields to these thermodynamic quantities.
The results turn out to be in a good agreement with the 
corresponding lattice data. In particular, the lattice-simulated peak of the interaction measure near the deconfinement 
critical temperature is reproduced. 
\end{abstract}

\maketitle

\section{Introduction}

How strong are the interactions in the quark-gluon plasma? In connection with  
RHIC heavy-ion experiments, this question has been posed~\cite{na} to understand the low shear viscosity 
necessary for a hydrodynamic description of the early phase of a heavy-ion collision~\cite{ro}. 
On the lattice, strong interactions in the QCD plasma up to temperatures about twice the 
deconfinement critical temperature are known from simulations of the interaction measure~\cite{f1,f2}.
Analytic models aiming at a description of these lattice data use both perturbative~\cite{bir} and 
nonperturbative~\cite{dps} techniques.
In the present paper, we revisit this problem by analytically calculating the pressure and the 
interaction measure $(\varepsilon-3p)/T^4$ of the gluon plasma. The main nonperturbative dynamical input 
of our calculation is the spatial confinement 
of valence gluons and the Polyakov loop, which are known from lattice simulations~\cite{bali} and~\cite{kk}, respectively.  Following 
previous works~\cite{dps, dp}, we use the term "valence gluons" emphasizing their confining interactions 
at large spatial distances. At small distances, valence gluons become quasi-free. Accordingly, the spatial 
Wilson loop 
of a valence gluon obeys the area law at large distances and goes to unity at small distances. 
Additionally, we model small spatial Wilson loops in such a way as to reproduce the known leading 
${\cal O}(g^2)$ perturbative contribution to the pressure of the gluon plasma~\cite{azh}. 
We manage to analytically calculate the 
world-line integral representing the free energy of a valence gluon at any distances. Furthermore, 
we account for the interaction of spatial valence gluons with temporal background fields through the Polyakov loop~\cite{dps}, and for stochastic background fields, which
provide spatial confinement of valence gluons. It turns out that our model is able to reproduce 
the lattice data on the pressure and the interaction measure of the gluon plasma rather well.

The world-line integral representing the one-loop free energy of a
valence gluon~\cite{dps, s} can be reduced to the Euler-Heisenberg effective Lagrangian in an auxiliary
{\it Abelian} field of a purely geometric origin. The weight factor associated with this field is constructed 
in such a way as to reproduce the Wilson loop in the original {\it non-Abelian} theory.  
We emphasize that we are not considering a YM theory in a
constant Abelian-type background field~\footnote{The reduction of the
world-line integral we are performing in this paper is similar to the
one used in Ref.~\cite{cond} in order to calculate the
condensates of a heavy quark through a tensor
{\it ansatz} for the minimal area~\cite{ta, ta21}.}.  
Another related approach to incorporate nonperturbative 
contributions to the one-loop effective
action~\cite{hrs} uses the two-point correlation
function of gluonic field strengths provided by the stochastic vacuum
model~\cite{ds}.

The paper is organized as follows. Section~II is devoted to the
calculation of the world-line integral for a valence gluon using a
parametrization for the minimal area in terms of the contour of
the spatial Wilson loop. In Section~III, we numerically evaluate the pressure and the 
interaction measure of the gluon plasma.
In Section~IV, we summarize our approach and the main results
of the paper. In Appendix~A, we evaluate the contributions of stochastic background fields to the 
pressure and the interaction measure.

\section{The set-up for the free-energy density}
Lattice simulations~\cite{bali, f1} in the
deconfinement phase ($T>T_c$) show that averages of large spatial Wilson loops
still exhibit an area law. In SU(3) YM theory, the critical temperature is
$T_c\simeq 270{\,}{\rm MeV}$~\cite{f1} (see e.g. Ref.~\cite{peter} for a review). 
This "magnetic" or "spatial" confinement~\cite{spconf} of
spatial Wilson-loop averages is the main established
nonperturbative phenomenon in the quark-gluon plasma. It 
does not contradict the genuine deconfinement of a static
quark-antiquark pair represented by averages of space-time Wilson
loops at $T>T_c$.
We follow the strategy of Refs.~\cite{dps, s, dp} 
to model the spatial confinement of the valence gluons $a_\mu^a$ by  
stochastic background fields $B_\mu^a$. 
With the {\it ansatz} $A_\mu^a=B_\mu^a+a_\mu^a$, the 
Euclidean YM partition function at finite temperature has the
following form:

$${\cal Z}(T)=\left<\int {\cal D}a_\mu^a({\bf x},t)\exp\left[-\frac{1}{4g^2}\int_0^\beta dx_4
\int_{V}^{}d^3x(F_{\mu\nu}^a[A])^2\right]\right>.$$
Here $\beta\equiv1/T$, $V$ is the three-dimensional volume, $F_{\mu\nu}^a[A]$ is the QCD field-strength tensor, 
and the averaging over the background $B_\mu^a$-fields is symbolized by the brackets 
$\left<\ldots\right>\equiv\left<\ldots\right>_B$. All the functional integrations 
imply periodic boundary conditions $a_\mu^a({\bf x},\beta)=a_\mu^a({\bf x},0)$, 
$B_\mu^a({\bf x},\beta)=B_\mu^a({\bf x},0)$.
Performing Gaussian integration over the valence gluons, one obtains the gluon determinant:

\begin{equation}
\label{int36}
{\cal Z}(T)=\left<\left\{\det\left[-(D_\mu^a[B])^2\right]\right\}^{-\frac12\cdot2(N_c^2-1)}\right>,
\end{equation}
with the covariant derivative 
$(D_\mu[B]f_\nu)^a=\partial_\mu f_\nu^a+f^{abc}B_\mu^bf_\nu^c$. 
Equation~(\ref{int36}) includes the 
color degrees of freedom of the valence gluons and their physical 
polarizations, $2(N_c^2-1)$. In the one-loop approximation for the $a_\mu^a$-field, it
can be simplified further: 

$${\cal Z}(T)=\left<\exp\left\{-(N_c^2-1){\rm Tr}{\,}\ln\left[-(D_\mu^a[B])^2\right]\right\}\right>
\simeq$$

\begin{equation}
\label{35tr}
\simeq\exp\left\{-(N_c^2-1)\left<{\rm Tr}{\,}\ln\left[-(D_\mu^a[B])^2\right]\right>\right\}.
\end{equation}  
The loop of fast field fluctuations receives infinitely many contributions
of the slow $B_\mu^a$-field.  In
Eq.~(\ref{35tr}), "Tr" includes the
trace "tr" over color indices and the functional trace over space-time
coordinates.

The free-energy density $F(T)$ is defined by the standard formula 

\begin{equation}
\label{st56}
\beta VF(T)=-\ln{\cal Z}(T).
\end{equation} 
The interactions of valence gluons with the background fields can be calculated through the
world-line method. Representing the trace of $\ln\left[-(D_\mu^a[B])^2\right]$
in terms of the Schwinger proper time $s$ as
${\rm Tr}{\,}\ln \hat O=-\int_0^\infty\frac{ds}{s}{\,}{\rm Tr}{\,}{\rm
e}^{-s\hat O}$, one has~\cite{oneloop}

\begin{equation}
\label{anfang}
F(T)=-(N_c^2-1)\cdot2\sum\limits_{n=1}^{\infty}\int_0^\infty\frac{ds}{s}
\int {\cal D}z_\mu {\rm e}^{-\frac14\int_0^sd\tau\dot z_\mu^2}
\left<W[z_\mu]\right>.
\end{equation}
Each world-line path $x_\mu(\tau)$ has a center $c_\mu=\frac1s\int_0^s d\tau x_\mu(\tau)$.
The integration over the positions of
the centers $c_\mu$ yields the factor $\beta V$ on the left-hand side of
Eq.~(\ref{st56}).  
The vector-function $z_\mu(\tau)$, $\tau\in[0,s]$, in Eq.~(\ref{anfang}) 
characterizes the shape of the contour of the Wilson loop relative to the center $c_\mu$.
The integration goes over world-line paths $z_\mu(\tau)=x_\mu(\tau)-c_\mu$ 
obeying the periodic boundary conditions 

\begin{equation}
\label{pbc}
z_4(s)=z_4(0)+\beta n,~~~~ {\bf z}(s)={\bf z}(0).
\end{equation}
Furthermore, the summation goes over the
winding number $n$ with a factor of 2 accounting for the modes with $n<0$.
The zero-temperature part of the
free-energy density corresponding to the zeroth winding mode, $n=0$, has
been subtracted~\cite{dps, HJP}. For simplicity, we use the standard
approximation which neglects the gluon spin term.
Equation~(\ref{anfang}) represents the main idea of our approach that the valence gluons,
as high-energy fluctuations of the gluonic field, propagate in a
low-energy stochastic background, which
enters Eq.~(\ref{anfang}) in the form of the Wilson-loop average

\begin{equation}
\label{w98}
\left<W[z_\mu]\right>\equiv\left<{\rm tr}{\,}{\cal P}\exp\left(i
\oint_{C}^{} dz_\mu B_\mu\right)\right>.
\end{equation} 
We use the notation $B_\mu\equiv B_\mu^at_{\rm adj}^a$ 
with $(t_{\rm adj}^a)^{bc}=-if^{abc}$. For brevity, we will 
call the average~(\ref{w98}) just "the Wilson loop"~\footnote{In our notation, the trace "tr" is normalized by dividing
over the trace of the unit matrix $\hat 1$ in the adjoint
representation, equal to $N_c^2-1$. Therefore, with this definition,
${\rm tr}{\,}\hat 1=1$.}. 
 
We will argue that the Wilson loop factorizes into a purely spatial Wilson loop and 
an averaged $n$-th power of a time-like Polyakov loop:

\begin{equation}
\label{f54}
\left<W[z_\mu]\right>\simeq\left<W[{\bf z}]\right>\prod\limits_{n=-\infty}^{+\infty}\left<L^n(T)\right>,
\end{equation}
with 

\begin{equation}
\label{spac}
\left<W[{\bf z}]\right>=\left<{\rm tr}{\,}{\cal
P}{\,}\exp\left(\frac{i}{2}\int_{\Sigma(C)} d\sigma_{jk}F_{jk}[B]
\right)\right>
\end{equation}
and 

$$\left<L^n(T)\right>=
\left<{\rm tr}{\,}{\cal T}\exp\left(in\int_0^{\beta} dz_4B_4\right)\right>.$$
In Eq.~(\ref{f54}), we have rewritten the spatial Wilson loop through the non-Abelian
Stokes' theorem. We have suppressed for simplicity the Schwinger strings, which connect the point where the 
field-strength tensor $F_{jk}^a$ is defined to a given reference point on  
an arbitrary surface $\Sigma(C)$ encircled by the contour $C$.  
The integration in Eq.~(\ref{spac}) goes
over the purely spatial part of that surface characterized by the differential
elements $d\sigma_{jk}$. By using the cumulant
expansion, one can express the Wilson loop $\left<W[z_\mu]\right>$ in
terms of the connected averages (cumulants) of chromo-magnetic and chromo-electric fields, 
$H_i^a=\frac12\varepsilon_{ijk}F_{jk}^a[B]$ and $E_i^a=iF_{4i}^a[B]$.
Lattice data~\cite{dmp} suggest that the cumulant-expansion
series can be truncated at the second term, since the amplitudes of
higher terms are significantly smaller. A general Wilson-loop
calculation  leads to three correlation functions

\begin{equation}
\left<E_i(x)E_k(x')\right>,~~ \left<H_i(x)H_k(x')\right>,~~ {\rm and}~~ \left<E_i(x)H_k(x')\right>.
\end{equation}
Following the lattice data~\cite{dmp}, we assume that the correlation
function $\left<E_i(x)E_k(x')\right>$ vanishes right after the
deconfinement phase transition. This fact ensures deconfinement.  The
same lattice data also indicate that the amplitude of the mixed
chromo-electric--chromo-magnetic correlation function
$\left<E_i(x)H_k(x')\right>$ is by an order of magnitude smaller than
the amplitude of the chromo-magnetic--chromo-magnetic correlation
function $\left<H_i(x)H_k(x')\right>$.  This property
leads to factorization, Eq.~(\ref{f54}). After factorization, the world-line integral over $z_4(\tau)$
becomes a one-dimensional world-line integral for a free
particle. Carrying it out, one arrives at the following expression
(cf. Ref.~\cite{dp}):

\begin{equation}
\label{0}
F(T)=-(N_c^2-1)\cdot2\sum\limits_{n=1}^{\infty}\int_0^\infty\frac{ds}{s}{\,}
\frac{{\rm e}^{-\frac{\beta^2n^2}{4s}}}{\sqrt{4\pi s}}\left<L^n(T)\right>
\int{\cal D}{\bf z}{\,}{\rm e}^{-\frac14\int_0^sd\tau\dot{\bf z}^2}
\left<W[{\bf z}]\right>.
\end{equation}

The dynamics of the strongly interacting gluon plasma is now encoded in
the Polyakov and spatial Wilson loop expectation values, which make the free energy nontrivial. 
An expectation value of the spatial Wilson loop $\left<W[{\bf z}]\right>$ depends on the area
$S\equiv S[{\bf z}]$ of the minimal surface defined by the
spatial contour ${\bf z}(s)$. At small spatial separations between valence gluons, 
$\left<W[{\bf z}]\right>\to 1$, that corresponds to a non-interacting Stefan-Boltzmann plasma. At large separations, a nonperturbative linear interaction between valence gluons emerges 
from the spatial area law, $\left<W[{\bf z}]\right>\to\exp[-\sigma(T)S]$. 
By "interaction" we mean the "potential" 
for the gluonic pair evolving in some spatial direction. The  
linear "potential" between valence gluons at large spatial separations can originate from their interaction with 
chromo-magnetic background fields $H_i^a$.

Furthermore, we mimic the leading ${\cal O}(g^2)$ perturbative contribution to the pressure~\cite{azh} by 
modeling perturbative interactions of valence gluons in terms of the spatial Wilson loop. The corresponding
short-distance {\it ansatz} for the renormalized spatial Wilson loop reads
$$\left<W[{\bf z}]\right>_{\rm p}={\rm e}^{-C(T)g^2(T)TL},$$
where $L=\int_0^s d\tau|\dot {\bf z}|$ is the length of the contour, and the prefactor 
$C(T)g^2(T)T$ is chosen in such a way that the leading ${\cal O}(g^2)$-correction to the 
Stefan-Boltzmann pressure is reproduced correctly. We associate the ${\cal O}(g^2)$-term in the pressure with perturbative interactions due to the massless-gluon exchange at short distances $\le{\cal O}\bigl(\frac1T\bigr)$.
These distances are smaller than ${\cal O}\bigl(\frac{1}{gT}\bigr)$
and ${\cal O}\bigl(\frac{1}{g^2T}\bigr)$, where the gluon mass terms in the propagator become effective
(cf. the recent paper~\cite{hp}). First, by comparing our result for the pressure with the known 
perturbative one~\cite{azh}, 
we will determine the limiting value $C$ of the function $C(T)$ at $T\gg T_c$. This limiting value alone 
is not sufficient, and in the second step, we 
will parametrize the functional form of $C(T)$.

Our short-distance {\it ansatz} for the spatial Wilson loop 
can be unified with the spatial-area law to a single formula
\begin{equation}
\label{twocases2}
\left<W[{\bf z}]\right>=\exp\left[-C(T)g^2(T)T\sqrt{S}-\sigma(T)S\right].
\end{equation}
For the short-distance part, we have estimated the length 
of the contour, $\int_0^s d\tau|\dot {\bf z}|$, as of order $\sqrt{S}$.
Note that, in this way, just one single equation~(\ref{twocases2}) 
reproduces both the small- and the large-distance regimes of the Wilson loop.
Indeed, the first term in the exponent, which describes perturbative interactions of valence gluons, dominates for sufficiently small areas, whereas the second term dominates for sufficiently large areas.

\begin{figure}
\epsfig{file=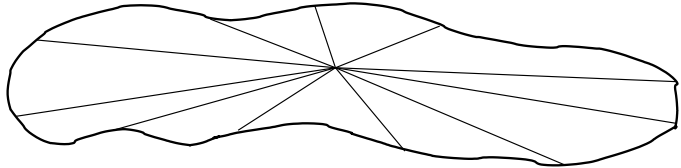, width=90mm}
\caption{\label{parasol} A view from above on a typical trajectory in the world-line integral, Eq.~(\ref{0}), depicted by the thick line, and the associated minimal surface
parametrized by Eq.~(\ref{SW}), depicted by the thin lines. 
The point where the pieces merge is the origin.}
\end{figure}

We implement now an {\it ansatz} for the minimal area $S$, which is 
crucial for the subsequent calculation:
\begin{equation}
\label{SW}
S\simeq\frac12\int_{0}^{s}d\tau|{\bf z}\times\dot{\bf z}|.
\end{equation}
It parametrizes a parasol-shaped surface made of thin segments, as depicted in Fig.~\ref{parasol}.
Note that, since $\int_0^s d\tau{\bf z}=0$, the point where the segments merge is the origin.
Thus, {\it ansatz}~(\ref{SW}) automatically selects from all cone-shaped surfaces having ${\bf z}(\tau)$ as a boundary
the one of the minimal area. We further approximate the integral in Eq.~(\ref{SW}) as follows:
\begin{equation}
\label{approxi}
S\simeq\sqrt{{\bf f}^2}
\end{equation}
with
$${\bf f}\equiv\frac12\int_{0}^{s}
d\tau({\bf z}\times\dot{\bf z}).$$
The left-hand side of
Eq.~(\ref{approxi}), given by Eq.~(\ref{SW}), can be larger than its right-hand side.  This
happens if, in the course of its evolution in the spatial directions,
the valence gluon performs backward and/or non-planar motions.  Should this happen, the
vector product $({\bf z}\times\dot{\bf z})$ changes its direction, and
the vector integral on the right-hand side of Eq.~(\ref{approxi})
receives mutually cancelling contributions. The non-backtracking approximation 
adopted in Eq.~(\ref{approxi}) is widely used in 
the context of minimal surfaces allowing analytic calculations of 
the associated world-line integrals (see e.g.~\cite{dp, cond, dks}).
We further proceed with calculating separately the perturbative part and the full 
free-energy density of the gluon plasma. 

\subsection{The perturbative contribution to the free-energy density}
At $T\gtrsim3T_c$, the Stefan-Boltzmann law along with the leading 
${\cal O}(g^2)$ perturbative correction~\cite{azh},
\begin{equation}
\label{kno}
F_{\rm p}(T)=-\frac{8\pi^2T^4}{45}\left[1-\frac{15}{16\pi^2}g^2(T)+{\cal O}(g^3)\right],
\end{equation}
reproduces more than 95\% of the lattice-simulated pressure of the gluon plasma~\cite{f1}. Therefore, at such temperatures,
one can retain only the first term in the exponent of Eq.~(\ref{twocases2}), where $C(T)$ should go to a constant 
$C$ with the increase of $T$. We determine the value of $C$ by calculating the perturbative contribution to the free-energy density with our {\it ansatz} for a small Wilson loop,
\begin{equation}
\label{Wspat}
\left<W[{\bf z}]\right>_{\rm p}={\rm e}^{-Cg^2(T)T({\bf f}^2)^{1/4}},
\end{equation}
and equating the result to the known expression, Eq.~(\ref{kno}).
The perturbative contribution to the free-energy density has the form
\begin{equation}
\label{Per}
F_{\rm p}(T)=-(N_c^2-1)\cdot2\sum\limits_{n=1}^{\infty}\int_0^\infty\frac{ds}{s}{\,}
\frac{{\rm e}^{-\frac{\beta^2n^2}{4s}}}{\sqrt{4\pi s}}
\int{\cal D}{\bf z}{\,}{\rm e}^{-\frac14\int_0^sd\tau\dot{\bf z}^2}
\left<W[{\bf z}]\right>_{\rm p},
\end{equation}
where we have used the leading-order perturbative result for the averaged Polyakov loop~\cite{GJ} 
\begin{equation}
\label{perTG}
\left<L(T)\right>_{\rm p}=1+{\cal O}(g^3).
\end{equation}
To calculate the world-line integral in Eq.~(\ref{Per}), we 
disentangle the fourth root in the exponent of Eq.~(\ref{Wspat}) by introducing two identical auxiliary integrations as follows:
$$
\left<W[{\bf z}]\right>_{\rm p}=
\frac{1}{\pi}\int_0^\infty\frac{d\lambda}{\sqrt{\lambda}}\int_0^\infty\frac{d\mu}{\sqrt{\mu}}
\exp\left[-\lambda-\mu-\frac{(Cg^2(T)T)^4{\bf f}^2}{64\lambda^2\mu}\right].$$
We further use the Hubbard-Stratonovich trick
\begin{equation}
\label{HS}
{\rm e}^{-A{\bf f}^2}=\frac{1}{(4\pi A)^{3/2}}\int d^3H{\rm e}^{-\frac{{\bf H}^2}{4A}+i{\bf H}{\bf f}},
\end{equation}
where $A>0$, and ${\bf H}$ is an auxiliary constant {\it Abelian} magnetic field. It yields 
$$
\left<W[{\bf z}]\right>_{\rm p}=$$
$$=\frac{64}{\pi^{5/2}}\frac{1}{(Cg^2(T)T)^6}\int_0^\infty d\lambda\lambda^{5/2}
{\rm e}^{-\lambda}\int_0^\infty d\mu\mu{\rm e}^{-\mu}\int d^3H\exp\left[-\frac{16\lambda^2\mu}{(Cg^2(T)T)^4}
{\bf H}^2+i{\bf H}{\bf f}\right].$$
This expression, plugged into the world-line integral, Eq.~(\ref{Per}), 
describes nothing but the proper-time evolution 
of a charged particle in the effective Abelian field ${\bf H}$,
governed by the Lorentz force $\propto (\dot{\bf z}\times{\bf H})$.  
The world-line integration, which is equivalent to the "summation" over all Landau levels, leads to the 
bosonic Euler-Heisenberg Lagrangian~\cite{oneloop}, namely 
\begin{equation}
\label{88}
\int{\cal D}{\bf z}{\,}{\rm e}^{-\frac14\int_{0}^{s}
d\tau\dot{\bf z}^2+i{\bf H}{\bf f}}=\frac{1}{(4\pi s)^{3/2}}{\,}\frac{Hs}{\sinh(Hs)}.
\end{equation}
The $\mu$-integration can be performed analytically. We further  
set $N_c=3$ and denote 
$$\xi\equiv Cg^2,$$ 
suppressing for brevity the temperature dependence of $g(T)$.
Rescaling $h\equiv H/\xi^2$, we obtain 
\begin{equation}
\label{fPert}
F_{\rm p}(T)
=-\frac{256T^4}{\pi^{7/2}}\xi^2\sum\limits_{n=1}^{\infty}
\int_0^\infty\frac{ds}{s^2}{\rm e}^{-\frac{n^2}{4s}}\int_0^\infty dh\frac{h^3}{\sinh(\xi^2hs)}
\int_0^\infty d\lambda\frac{\lambda^{5/2}{\rm e}^{-\lambda}}{(16\lambda^2h^2+1)^2}.
\end{equation}
To the leading order of the $\xi$-expansion, $\sinh(\xi^2hs)\simeq\xi^2hs$, and one can use the value of the 
integral 
$$\int_0^\infty dhh^2\int_0^\infty d\lambda\frac{\lambda^{5/2}{\rm e}^{-\lambda}}{(16\lambda^2h^2+1)^2}=
\frac{\pi^{3/2}}{256}$$
to recover the Stefan-Boltzmann result:
\begin{equation}
\label{sb12}
F_{\rm p}(T)\simeq-\frac{T^4}{\pi^2}\sum\limits_{n=1}^{\infty}\int_0^\infty\frac{ds}{s^3}{\rm e}^{-\frac{n^2}{4s}}=
-\frac{8\pi^2T^4}{45}.
\end{equation}
The next term of the $\xi$-expansion of Eq.~(\ref{fPert}) should correspond to the $g^2$-term in Eq.~(\ref{kno}).
To derive it, we should approximate $\sinh(\xi^2hs)\simeq\xi^2hs[1+(\xi^2hs)^2/6]$, that yields 
$$F_{\rm p}(T)\simeq-\frac{256T^4}{\pi^{7/2}}\sum\limits_{n=1}^{\infty}
\int_0^\infty\frac{ds}{s^3}{\rm e}^{-\frac{n^2}{4s}}
\int_0^\infty d\lambda\lambda^{5/2}{\rm e}^{-\lambda}\int_0^\infty dh\frac{h^2}{(16\lambda^2h^2+1)^2}
\cdot\frac{1}{1+(\xi^2hs)^2/6}.$$
The $h$-integration in this formula can be performed analytically, so that we obtain 
$$F_{\rm p}(T)\simeq-\frac{16T^4}{\pi^{5/2}}\sum\limits_{n=1}^{\infty}
\int_0^\infty\frac{ds}{s^3}{\rm e}^{-\frac{n^2}{4s}}\int_0^\infty d\lambda
\frac{\lambda^{3/2}{\rm e}^{-\lambda}}{(4\lambda+\xi^2s/\sqrt{6})^2}.$$
To the order ${\cal O}(\xi^0)$, this formula recovers again the Stefan-Boltzmann free-energy density, Eq.~(\ref{sb12}).
The ${\cal O}(\xi^1)$-term of interest can be obtained by approximating the sum over winding modes 
by the first two terms:
$$\int_0^\infty\frac{ds}{s^3}\left({\rm e}^{-\frac{1}{4s}}+{\rm e}^{-\frac1s}\right)
\int_0^\infty d\lambda
\frac{\lambda^{3/2}{\rm e}^{-\lambda}}{(4\lambda+\xi^2s/\sqrt{6})^2}=
\frac{17\sqrt{\pi}}{16}-\frac{27\pi^{3/2}}{128\cdot 6^{1/4}}\cdot\xi+{\cal O}(\xi^2).$$
This yields the desired free-energy density 
\begin{equation}
\label{ddk}
F_{\rm p}(T)\simeq-T^4\left(\frac{17}{\pi^2}-\frac{27}{8\pi\cdot 6^{1/4}}\cdot Cg^2(T)\right),
\end{equation}
which can be compared with the known result, Eq.~(\ref{kno}). Comparing the ${\cal O}(g^0)$-terms,
$\frac{8\pi^2}{45}\simeq1.75$ and $\frac{17}{\pi^2}\simeq1.72$, we conclude that the approximation of the 
full sum over winding modes by the $(n=1)$- and the $(n=2)$-terms is indeed very good. 
Comparing the ${\cal O}(g^2)$-terms, we finally obtain:
\begin{equation}
\label{CC}
C=\frac{8\pi}{27\cdot 6^{3/4}}\simeq0.24.
\end{equation}
We set the function $C(T)$ to this constant value at the highest temperature of $4.54T_c$ achievable by the 
lattice simulations. With the decrease of temperature, perturbation theory becomes less and less
applicable. We impose this fade of perturbation theory by the gradual nullification of the function $C(T)$.
The full free-energy density, corresponding to the spatial Wilson loop~(\ref{twocases2}), will be 
calculated in the next Subsection.

\subsection{The full free-energy density}
To calculate the world-line integral with the full Wilson loop, Eq.~(\ref{twocases2}), we first transform 
$\sqrt{S}$ in the exponent to $S$ by introducing an auxiliary integration as
$$\left<W[{\bf z}]\right>={\rm e}^{-\gamma\sqrt{S}-\sigma S}=\int_0^\infty\frac{d\lambda}{\sqrt{\pi\lambda}}
\exp\left[-\lambda-\left(\frac{\gamma^2}{4\lambda}+\sigma\right)S\right],$$
where 
$$\gamma\equiv C(T)g^2(T)T,$$ 
$\sigma\equiv\sigma(T)$. 
By introducing one more such integration, we further transform $S$ to $S^2$:
$$\left<W[{\bf z}]\right>=\int_0^\infty\frac{d\lambda}{\sqrt{\pi\lambda}}\int_0^\infty\frac{d\mu}{\sqrt{\pi\mu}}
\exp\left[-\lambda-\mu-\frac{1}{4\mu}\left(\frac{\gamma^2}{4\lambda}+\sigma\right)^2S^2\right].$$
We can now apply the Hubbard-Stratonovich trick, Eq.~(\ref{HS}), and integrate over $\mu$ analytically.
With the use of Eq.~(\ref{approxi}), this yields
\begin{equation}
\label{W97}
\left<W[{\bf z}]\right>=\frac{1}{\pi^{5/2}}\int_0^\infty\frac{d\lambda}{\sqrt{\lambda}}{\rm e}^{-\lambda}
\int d^3H\frac{\frac{\gamma^2}{4\lambda}+\sigma}{\left[{\bf H}^2+\left(\frac{\gamma^2}{4\lambda}+
\sigma\right)^2\right]^2}{\,}{\rm e}^{i{\bf H}{\bf f}}.
\end{equation}
The world-line integration can again be performed by using Eq.~(\ref{88}), that finally 
yields the full free-energy density
$$F(T)=-\frac{4}{\pi^{7/2}}\times$$
\begin{equation}
\label{fin}
\times\sum\limits_{n=1}^{\infty}\left<L^n(T)\right>\int_0^\infty\frac{ds}{s^2}
{\rm e}^{-\frac{\beta^2n^2}{4s}}\int_0^\infty\frac{d\lambda}{\sqrt{\lambda}}\left(\frac{\gamma^2}{4\lambda}+
\sigma\right){\rm e}^{-\lambda}\int_0^\infty dH\frac{H^3/\sinh(Hs)}{\left[H^2+\left(\frac{\gamma^2}{4\lambda}+
\sigma\right)^2\right]^2}.
\end{equation}
The remaining {\it ordinary} integrations over $H$, $\lambda$, and $s$ will be performed numerically in the next Section.

\section{Numerical evaluation}

We fix now the temperature dependences of various quantities entering the calculation. 
We adopt the perturbative two-loop running coupling at finite temperature~\cite{f1}
\begin{equation}
\label{g2}
g^{-2}(T)=2b_0\ln\frac{T}{\Lambda}+\frac{b_1}{b_0}\ln\left(2\ln\frac{T}{\Lambda}\right),~ {\rm where}~
b_0=\frac{11N_c}{48\pi^2},~ b_1=\frac{34}{3}\left(\frac{N_c}{16\pi^2}\right)^2,
\end{equation}
and $N_c=3$ for the case under study. Furthermore, we assume the value
$\Lambda=0.104T_c$~\cite{f1}. 

\begin{figure}
\psfrag{P}{\Large{$\left<L_{\rm f}(T)\right>$}}
\epsfig{file=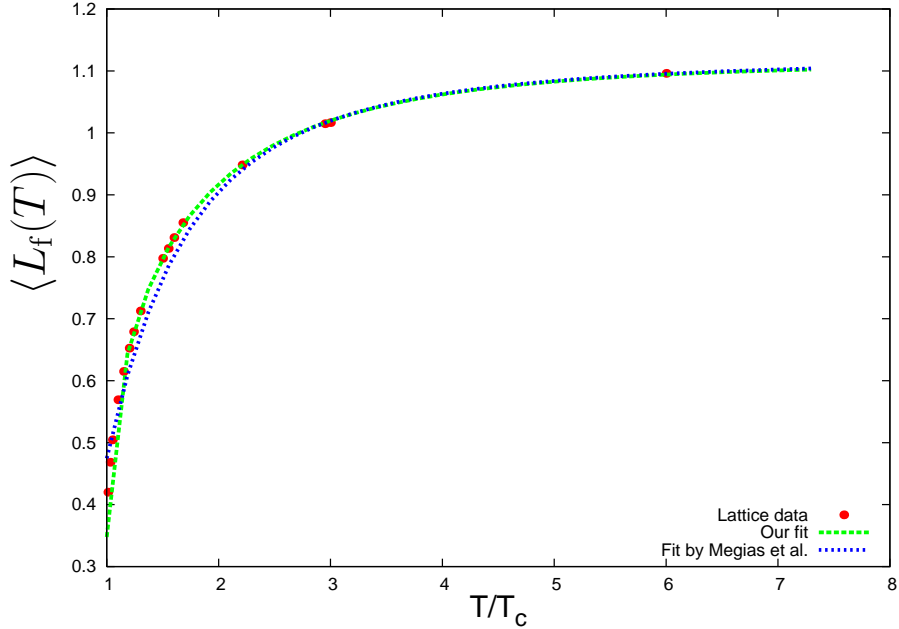, width=120mm}
\caption{\label{fig1}The averaged Polyakov loop $\left<L_{\rm f}(T)\right>$ according to Ref.~\cite{kk} 
(courtesy of P.~Petreczky) and the fitting curves according to Eqs.~(\ref{69}) and~(\ref{669}).}
\end{figure}

At temperatures smaller than 
the temperature of dimensional reduction, $T<T_{*}$, all dimensionful quantities can be approximated
by their zero-temperature values~\cite{aga}. We use the value $T_{*}=2T_c$, as suggested by the same 
lattice data~\cite{f1}.
The temperature-dependent string tension in the adjoint representation of interest reads
\begin{equation}
\label{siG}
\sigma(T)=\frac94\sigma_0\cdot\left\{\begin{array}{rcl}1~~ {\rm at}~~ T_c<T<T_{*},\\
\left[\frac{g^2(T)}{g^2(T_{*})}\cdot\frac{T}{T_{*}}\right]^2~~ {\rm at}~~ T>T_{*},
\end{array}\right.
\end{equation}
where $\sigma_0=(440{\,}{\rm MeV})^2$ is the zero-temperature value of the  
string tension in the fundamental representation.
The coefficient 9/4 stems from the Casimir-scaling
hypothesis for the string tensions in various representations.
Casimir scaling has proven in Ref.~\cite{GHK} to be a good approximation also for the averaged 
Polyakov loop in the adjoint representation, $\left<L(T)\right>$. That is, this average can be approximated as 
$\left<L(T)\right>\simeq\left<L_{\rm f}(T)\right>^{9/4}$, where $\left<L_{\rm f}(T)\right>$ is the averaged 
Polyakov loop in the fundamental representation, which has been simulated on 
the lattice in Ref.~\cite{kk}. Since the $(N_\tau=4)$- and $(N_\tau=8)$-data from Ref.~\cite{kk} follow the 
same pattern, we fit them both by the four-parameter function, which we use in the subsequent calculation:

\begin{equation}
\label{69}
\left<L_{\rm f}(T)\right>=1.11-\frac{1.53}{1+\exp\left[1.92\left(\frac{T}{T_c}-1\right)^{0.51}\right]}.
\end{equation}
In Fig.~\ref{fig1}, we plot both the lattice data on the averaged Polyakov loop $\left<L_{\rm f}(T)\right>$
and the results of the fit. For comparison, in the same Fig.~\ref{fig1}, we present the results of 
a simpler fit for the $(N_\tau=8)$-data alone~\cite{M}
\begin{equation}
\label{669}
\left<L_{\rm f}(T)\right>=\exp\left\{-\frac12\left[-0.23+1.72(T_c/T)^2\right]\right\}
\end{equation}
and observe a bit less good agreement. Note that, at $T\gg T_c$, our fit for $\left<L_{\rm f}(T)\right>$ and the fit of Ref.~\cite{M} approach unity from above, in accordance with the positivity of the ${\cal O}(g^3)$-correction
in Eq.~(\ref{perTG})~\cite{GJ, hp}.
Furthermore, to estimate the effect of the second winding mode, we use 
for the averaged square of the Polyakov loop in the adjoint representation the following approximation: $\left<L^2(T)\right>\simeq\left<L(T)\right>^2$. Note that the 
freedom in choice of the renormalization scheme results in the 
multiplication of the renormalized Polyakov loop by a temperature-dependent factor.
Such a multiplication is equivalent to the adding of some constant $A$ to 
the static quark-antiquark potential. We use here the values of the 
renormalized Polyakov loop in the fundamental representation from Ref.~\cite{kk}. In that paper, 
$A$ is fixed to 0 by demanding the static potential at $T=0$ to have the form
$V(r)=\sigma_0 r - \pi/(12 r)$ at $r>r_0$, where the value of $\sigma_0$ is given right after Eq.~(\ref{siG}), 
and $r_0\simeq 0.5{\,}{\rm fm}$ is the so-called Sommer scale~\cite{som}.
Finally, to minimize the number of free parameters, we assume 
that the function $C(T)$ rises linearly from 0 to the derived value~(\ref{CC}) when $T$ grows from $T_{*}$ to
$4.54T_c$, that is the maximal temperature at which the lattice values~\cite{f1} for the pressure are available.
Explicitly, the assumed function $C(T)$ has the form
\begin{equation}
\label{C1}
C(T)=
\left\{\begin{array}{rcl}0~~~~~~~~~~~~~ {\rm at}~~ T_c<T<T_{*},\\
0.24\cdot \frac{T-T_{*}}{4.54T_c-T_{*}}~~ {\rm at}~~ T_{*}<T<4.54T_c.
\end{array}\right.
\end{equation}

Using the above parametrizations, we calculate numerically the pressure of valence gluons 
$$p_{\rm val}(T)=-F(T),$$ 
with their full free-energy density given by Eq.~(\ref{fin}). This further yields 
their contribution to the interaction measure 
$$(\varepsilon-3p)_{\rm val}=T\frac{\partial p_{\rm val}}{\partial T}-4p_{\rm val}.$$
Additionally, we take into account stochastic background fields, which provide 
spatial confinement of valence gluons in our model. Their contribution $(\varepsilon-3p)_{\rm stoch}$
to the interaction measure is evaluated in Appendix~A. We find that $(\varepsilon-3p)_{\rm stoch}$
is nonvanishing and negative definite 
at temperatures $T>T_{*}$, where its absolute value amounts to a few percent of $(\varepsilon-3p)_{\rm val}$.

\begin{figure}
\psfrag{Q}{\Large{$\frac{\varepsilon-3p}{T^4}$}}
\epsfig{file=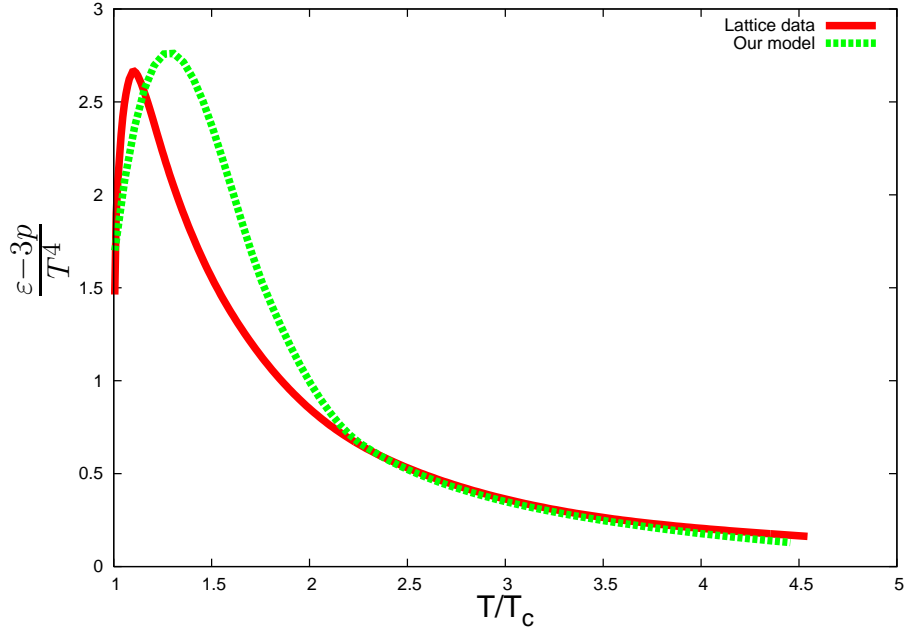, width=120mm}
\caption{\label{WW}Lattice data (full drawn curve)~\cite{f1} on the interaction measure $\frac{\varepsilon-3p}{T^4}$ (courtesy of F.~Karsch) compared to the results of our model (dashed curve).}
\end{figure}

\begin{figure}
\psfrag{Q}{\Large{$\frac{p}{T^4}$}}
\epsfig{file=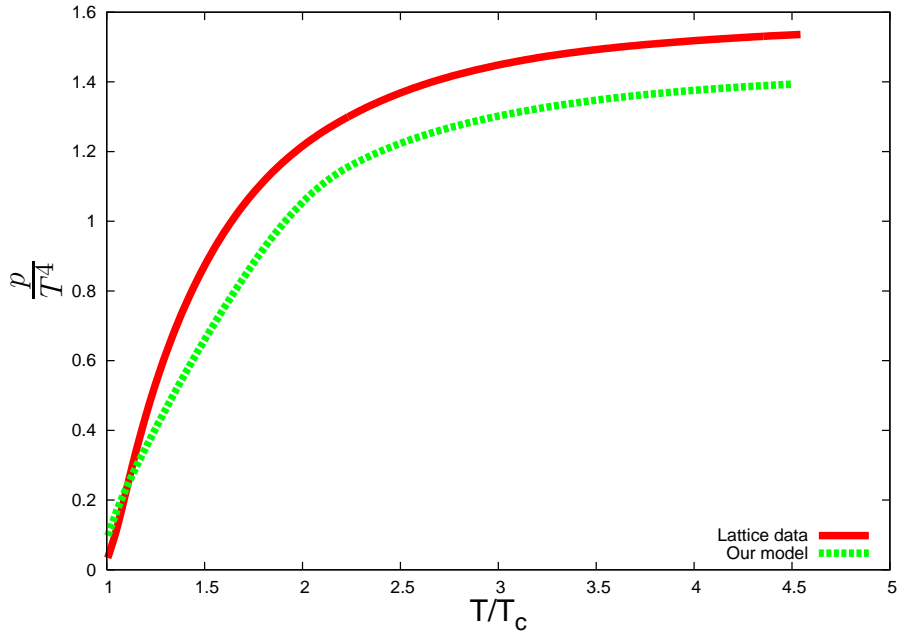, width=120mm}
\caption{\label{pR}Lattice data (full drawn curve)~\cite{f1} on the ratio $p/T^4$ (courtesy of F.~Karsch) compared to the results of our model (dashed curve).}
\end{figure}

In Fig.~\ref{WW}, we plot the full interaction measure 
$$\frac{\varepsilon-3p}{T^4}=\frac{(\varepsilon-3p)_{\rm val}+(\varepsilon-3p)_{\rm stoch}}{T^4},$$
and compare it with the lattice results from Ref.~\cite{f1}. In Fig.~\ref{pR}, we do the same for the 
full pressure,
$$\frac{p}{T^4}=\frac{p_{\rm val}+p_{\rm stoch}}{T^4}.$$
Given all the approximations adopted, 
one can conclude that, for both the interaction measure and the pressure, the agreements with the corresponding lattice results are rather good.

\begin{figure}
\psfrag{Q}{\Large{$\frac{p}{T^4}$}}
\epsfig{file=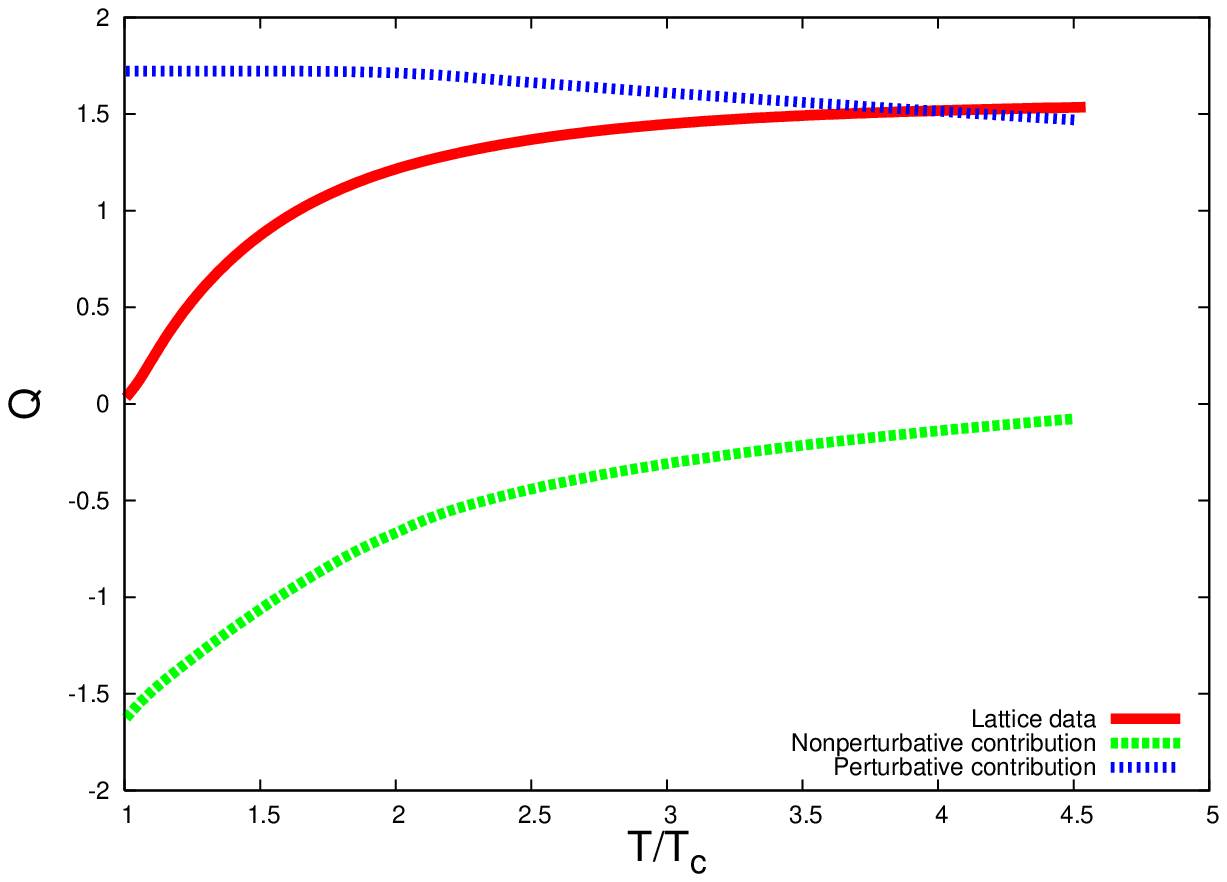, width=120mm}
\caption{\label{uu}Separately shown are the nonperturbative and the perturbative contributions to the ratio $p/T^4$. The 
lattice data, as in Fig.~\ref{pR}, are shown once again for comparison.}
\end{figure}

Finally, one can define separately the perturbative and the 
nonperturbative contributions to the pressure. The perturbative contribution is given by Eq.~(\ref{ddk})
with the constant $C$ replaced by the full function $C(T)$ given by Eq.~(\ref{C1}). Explicitly, this 
contribution reads~\footnote{Note that, at temperatures $T_c<T<T_{*}$, the second term in this expression vanishes, and $p_{\rm pert}$ goes over to the Stefan-Boltzmann value approximated by the first two winding modes.} 
$$\frac{p_{\rm pert}}{T^4}=\frac{17}{\pi^2}-\frac{27}{8\pi\cdot 6^{1/4}}\cdot C(T)g^2(T).$$
The nonperturbative contribution is accordingly defined as a difference 
$$\frac{p_{\rm nonpert}}{T^4}=\frac{p-p_{\rm pert}}{T^4}.$$
It includes the pressure produced by stochastic background fields and the nonperturbative part of the 
pressure produced by valence gluons.
In Fig.~\ref{uu}, we plot $p_{\rm pert}/T^4$, $p_{\rm nonpert}/T^4$, and once again 
for comparison the lattice data on $p/T^4$.

\section{Summary and concluding remarks}
The aim of the present paper has been an analytic calculation of the pressure $p(T)$ and the interaction measure 
$(\varepsilon-3p)/T^4$ of the gluon plasma in SU(3) YM theory. Our main phenomenological input is confinement 
of valence gluons at large {\it spatial} separations and their perturbative interaction at small separations.
The corresponding form of a spatial Wilson loop, Eq.~(\ref{twocases2}), accounts for these interactions.

One can prove numerically that, 
at temperatures $T\gtrsim3T_c$, the lattice-simulated pressure of the gluon plasma 
is by more than 95\% described by the Stefan-Boltzmann law along with the perturbative interaction between 
valence gluons, Eq.~(\ref{kno}). This fact enables us to fix the limiting high-temperature value $C$ of the function $C(T)$, which is used to parametrize the perturbative interaction in terms of spatial Wilson loops. The value of $C$, Eq.~(\ref{CC}),
is obtained by analytically calculating the world-line integral with such a Wilson loop and comparing the obtained
free-energy density with the known result~(\ref{kno}).

At high temperatures, $T>2T_c$, valence gluons in our model interact both linearly at large spatial 
separations and Coulomb-like at small separations, while at $T<2T_c$ only the large-distance 
linear part remains.
With these interactions taken into account, we manage to analytically 
calculate the world-line integral, and obtain the contribution of spatial dynamics to the pressure 
at any temperature, Eq.~(\ref{fin}).
Additionally, we account also for the interaction of valence gluons with the $A_4^a$-gluons
via the Polyakov loop, which is taken as a fit to the lattice data~\cite{kk, GHK}. 
The critical behavior of the Polyakov loop is crucial for the nullification of the pressure at $T=T_c$
and a peak of the interaction measure near $T_c$ (cf. Refs.~\cite{dps, mom}). Furthermore, 
stochastic background fields, that can be responsible for the spatial confinement of valence gluons, produce 
their own contributions to the interaction measure and the pressure, which are also taken into account 
[cf. Eqs.~(\ref{3uu}) and (\ref{4uu})].
In Figs.~\ref{WW} and \ref{pR}, we compare our results for these two thermodynamic quantities of the gluon plasma 
with the corresponding lattice data.
The observed reasonable 
agreement suggests that spatially confined valence gluons, along with the stochastic 
background fields, could indeed be the right degrees of 
freedom for the description of the gluon-plasma thermodynamics.

Last but not least, we mention that, on the technical side, our paper advanced to analytically calculate
the world-line integral with the minimal-area surface for each trajectory involved in the integration. 
Of course, it became possible
only by virtue of a certain parametrization of such a surface in terms of the corresponding trajectory. 
Namely, we first assume that the surface has a parasol-shaped form, so that its area is
$S\simeq\frac12\int_{0}^{s}d\tau|{\bf z}\times\dot{\bf z}|$, and use the approximation of 
non-backtracking planar trajectories $S\simeq\frac12\left|\int_{0}^{s} d\tau({\bf z}\times\dot{\bf z})\right|$. With the 
Hubbard-Stratonovich transformation~(\ref{HS}), furthermore it is possible to single out the trajectory-dependence 
in the form $\exp\bigl[\frac{i}{2}{\bf H}\int_0^s d\tau({\bf z}\times\dot{\bf z})\bigr]$, where 
${\bf H}$ is an {\it auxiliary constant Abelian} magnetic field. For such a field, the world-line integral 
becomes the known Euler-Heisenberg Lagrangian, Eq.~(\ref{88}). It looks promising to 
generalize this parametrization to the 4D case, that amounts to additionally introducing an effective
Abelian {\it electric} field. Such an approach would enable one to systematically account for effects of confinement 
in various Feynman diagrams (cf. Ref.~\cite{dp}). Work in this direction is in progress.

Finally, let us mention other recent studies, where the same thermodynamic quantities have been 
calculated either by analytic methods, perturbatively in~\cite{ML} and nonperturbatively in~\cite{yus}, or on the lattice~\cite{max, panero}. In particular, Ref.~\cite{panero} generalizing the interaction measure obtained in Ref.~\cite{f1} to $N_c>3$ states that $\varepsilon-3p$ is approximately independent of $N_c$ for $N_c\le 6$. 
This observation, once receiving confirmation by further lattice measurements, deserves theoretical interpretation.

\section*{Acknowledgements} 
We are grateful for usefuls discussions and correspondences to O.~Andreev, J.-P. Blaizot,
D.~Gr\"unewald, E.-M.~Ilgenfritz, O.~Kaczmarek, F.~Karsch, M.~Laine, E.~Megias, M.~Panero, and Yu.A.~Simonov.
We also thank F.~Karsch and P.~Petreczky for
providing the details of the lattice data. The work of D.A. has been supported by 
the Helmholtz Alliance Program of the Helmholtz Association, contract HA-216
"Extremes of Density and Temperature: Cosmic Matter in the Laboratory", at the initial stage, and by 
the German Research Foundation (DFG), contract Sh~92/2-1, at the final stage.

\section*{Appendix A. Contributions of stochastic background fields to the pressure and the interaction measure}
The contribution of stochastic background fields to the interaction measure 
can be calculated directly, by the known formula, which assumes no valence gluons~\cite{leu}
\begin{equation}
\label{nval}
(\varepsilon-3p)_{\rm no{\,}val.{\,}gl.}=\frac{-b}{32\pi^2}\left[\left<(gH_i^a)^2\right>_T-
\left<(gH_i^a)^2\right>_0\right]
\end{equation}
with $b=\frac{11}{3}N_c+\frac{34}{3}\frac{N_c^2}{16\pi^2}$ in the two-loop approximation adopted 
[cf. Eq.~(\ref{g2})], and $N_c=3$. 
The contribution of the 
chromo-magnetic condensate to the trace anomaly in the vacuum is 
subtracted. It amounts to a half of the full contribution, while the other half is represented 
by the chromo-electric gluon condensate $\left<(gE_i^a)^2\right>_T$, whose  
``evaporation'' at $T=T_c$ leads to deconfinement~\cite{s, ds, dmp}. Thus,
$\left<(gH_i^a)^2\right>_0=\frac12\left<(gF_{\mu\nu}^a)^2\right>_0$, where 
the value of the full 
condensate consistent with the known values of the string tension and of 
the vacuum correlation length~\cite{dmp} is 
$\left<(gF_{\mu\nu}^a)^2\right>_0=3.55{\,}{\rm GeV}^4$~\cite{dp, sFP}.
The temperature dependence of the chromo-magnetic condensate can be 
modeled by a formula similar to Eq.~(\ref{siG}),
\begin{equation}
\label{2uu}
\left<(gH_i^a)^2\right>_T=\left<(gH_i^a)^2\right>_0\cdot
\left\{\begin{array}{rcl}1~~ {\rm at}~~ T_c<T<T_{*},\\
\left[\frac{g^2(T)}{g^2(T_{*})}\cdot\frac{T}{T_{*}}\right]^4~~ {\rm at}~~ T>T_{*}.
\end{array}\right.
\end{equation}
According to Eqs.~(\ref{nval}) and (\ref{2uu}), stochastic background fields alone produce no contribution to the 
interaction measure at $T_c<T<T_{*}$, due to the constancy of $\left<(gH_i^a)^2\right>_T$ 
at these temperatures~\footnote{We disregard the exponentially small corrections to the constant values 
of $\sigma(T)$ and $\left<(gH_i^a)^2\right>_T$ at $T_c<T<T_{*}$, whose explicit form can be found in Ref.~\cite{aga}.}.
At $T>T_{*}$, this contribution starts appearing, but at the highest temperature under consideration, 
$T=4.54T_c$, it must vanish again, as we assume that at this temperature all thermodynamic quantities are 
completely saturated by perturbative interactions of valence gluons. Since the increase of the perturbative contribution 
is given by the function~(\ref{C1}), 
the simultaneous decrease of the nonperturbative contribution should be described by the function $[C-C(T)]/C$, 
where $C=C(4.54T_c)\simeq0.24$, Eq.~(\ref{CC}).
Hence, the contribution of stochastic background fields to the interaction measure can be parametrized as 
\begin{equation}
\label{3uu}
(\varepsilon-3p)_{\rm stoch}=\frac{C-C(T)}{C}\cdot\frac{(-b)}{32\pi^2}\left[\left<(gH_i^a)^2\right>_T-
\left<(gH_i^a)^2\right>_0\right].
\end{equation}
The contribution to the pressure produced by stochastic background fields, $p_{\rm stoch}$,
can be obtained by integrating the formula
$$T\frac{\partial}{\partial T}\frac{p_{\rm stoch}}{T^4}=\frac{(\varepsilon-3p)_{\rm stoch}}{T^4}$$
from $T=T_{*}$, below which $(\varepsilon-3p)_{\rm stoch}=0$, on. This contribution thus reads
\begin{equation}
\label{4uu}
\frac{p_{\rm stoch}(T)}{T^4}=\int_{T_{*}}^{T}dT'\frac{(\varepsilon-3p)_{\rm stoch}(T')}{T'^5}.
\end{equation}
As follows from Eqs.~(\ref{2uu})-(\ref{4uu}), both $(\varepsilon-3p)_{\rm stoch}$ and $p_{\rm stoch}$
are negative definite, leading to the decreases of the full $(\varepsilon-3p)$ and $p$ by a 
few percent at $T_{*}<T<4.54T_c$.

\end{document}